\documentclass{ws-p8-50x6-00}

\def\Journal#1#2#3#4{{#1} {#2}, #3 (#4)}

\def\NPB{{\em Nucl. Phys.} B}
\def\PLB{{\em Phys. Lett.}  B}

\def\ZPC{{\em Z. Phys.} C}

\def\be{\begin{equation}}
\def\ee{\end{equation}}
\def\bea{\begin{eqnarray}}
\def\eea{\end{eqnarray}}

\def\ifmath#1{\relax\ifmmode #1\else $#1$\fi}%
\def\rd{\ifmath{{\mathrm{d}}}}
\def\rt{\ifmath{{\mathrm{t}}}}
\def\rT{\ifmath{{\mathrm{T}}}}
\def\rZ{\ifmath{{\mathrm{Z}}}}

\def\data{\ifmath{{\mathrm{data}}}}
\def\lon{\ifmath{{\mathrm{long}}}}
\def\side{\ifmath{{\mathrm{side}}}}
\def\out{\ifmath{{\mathrm{out}}}}
\def\true{\ifmath{{\mathrm{true}}}}
\def\reference{\ifmath{{\mathrm{reference}}}}
\def\MonteCarlo{\ifmath{{\mathrm{MonteCarlo}}}}
\def\vec#1{{\mbox{\bf #1}}}

\begin{document}
\title{MULTIPARTICLE CORRELATIONS AT LEP1}
\author{OXANA SMIRNOVA}
\address{Particle Physics, Institute of
    Physics, Lund University, P.O.~Box 118, 22100 Lund,
    Sweden (on leave from JINR, Dubna, 141980
      Russia). \\E-mail: oxana.smirnova@quark.lu.se}
\maketitle
\abstracts{Bidding farewell to the LEP accelerator, we
  acknowledge that an essential part of its legacy is the immense
  amount of data collected during 5 years of the LEP1 
   stage, at the $\rZ^0$ peak. This set of data allows for detailed
   studies of many phenomena, and particularly, of the final state
   interactions between hadrons. Presented here is a review 
   of the most recent analyses in this area.
}

\section{Introduction}
This talk presents an overview of latest results in multiparticle
correlation studies in hadronic decays of $\rZ^0$ by LEP experiments.
Unless stated otherwise, it concerns pairs of identically charged particles.
Analyses to be reviewed include the observation
of a pion source elongation, studies of correlations between kaons
and between $\Lambda$ baryon pairs, and the transverse mass
dependence of the two-boson correlation function parameters.

Multiparticle correlations are best described in terms of the
four-momentum difference, $Q$, defined for two particles with
four-momenta $(E_1,\vec{p}_1)$ and $(E_2,\vec{p}_2)$ as 
$Q=\sqrt{(\vec{p}_1-\vec{p}_2)^2-(E_1-E_2)^2}$.
A two-particle correlation function is most commonly parametrized by
a Gaussian shape:
\begin{equation}
C_2(Q)=1+\lambda\cdot e^{-Q^2 R^2}\ ,
\label{eq:c2def}
\end{equation}
where $R$ is usually interpreted as the geometrical
source size, and $\lambda$ reflects the strength of the effect.
Presence of Bose-Einstein correlations (BEC) should lead to enhanced
production of boson pairs with small $Q$ (positive $\lambda$), while
Fermi-Dirac correlations (FDC) suppress production of fermion pairs in
the region of small $Q$ (negative $\lambda$). Extra terms are
occasionally used to account for the experimental background.

The transverse mass, $m_\rT$, of two particles is defined as
\be
m_\rT=\left(\sqrt{m_1^2+p_{\rT,1}^2}+\sqrt{m_2^2+p_{\rT,2}^2}\right)/2\ ,
\label{eq:mtdef}
\ee
where $m_1$, $m_2$ are the particle masses, and $p_{\rT,1}$, $p_{\rT,2}$ are
their transverse momenta with respect to the axis of the process.

For multidimensional analysis in components of $Q$, it is convenient
to introduce the Longitudinal Centre of Mass System (LCMS),\cite{lcms}
in which $Q$ is resolved into a longitudinal component $Q_{\lon}$
(aligned with a process axis), and two transverse ones: $Q_{\rt,\out}$
and $Q_{\rt,\side}$. The three-dimensional representation of two-particle
correlations in LCMS is then
\be
C_2(Q_{\rt,\out},Q_{\rt,\side},Q_{\lon})=1+\lambda\cdot
e^{-Q^2_{\rt,\out}R^2_{\rt,\out}
  -Q^2_{\rt,\side}R^2_{\rt,\side}-Q^2_{\lon}R^2_{\lon}}\ .
\label{eq:c23dim}
\ee

Experimentally, the correlation-caused enhancement (or
depletion) is measured in the ratio of
\be
R(Q)=\left((\rd N/\rd Q)_{\data}\right)/\left((\rd N/\rd Q)_{\reference}\right)\ ,
\label{ratio1}
\ee
where $N$ is the number of particle pairs. The choice of the reference
sample $(\rd N/\rd Q)_{\reference}$, i.e., the one without the
correlations in question, is very important. Possible reference
distributions can be made by: a) pairing unlike-charge particles
(often denoted as ``$+-$''); b)
``event-mixing'', i.e., combining like-charge pairs from particles
belonging to different events; c) using Monte Carlo generated events
without correlations. None of the methods is absolutely preferred, as
each of them has 
specific shortcomings. It will be shown below that results might depend
strongly on the reference sample choice. To compensate for the
introduced side-effects, a double-ratio is often used to measure the
``true'' correlation: $R^{\true}(Q)=R_{\data}(Q)/R_{\MonteCarlo}(Q)$.
\section{Pion source elongation}
\label{sec:2dim}
\begin{figure}[hb]
\begin{center}
\includegraphics[width=4cm]{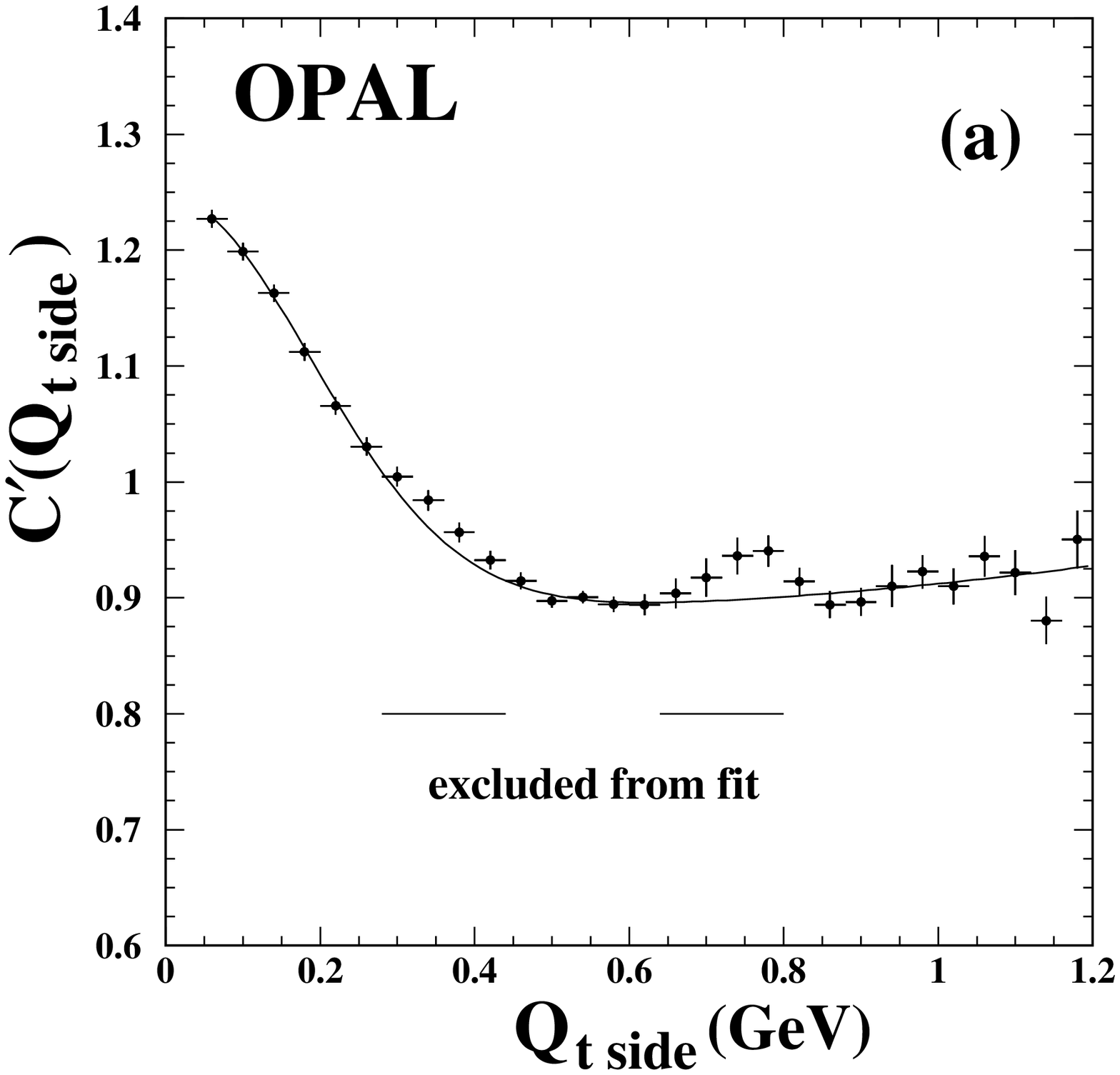}
\includegraphics[width=4cm]{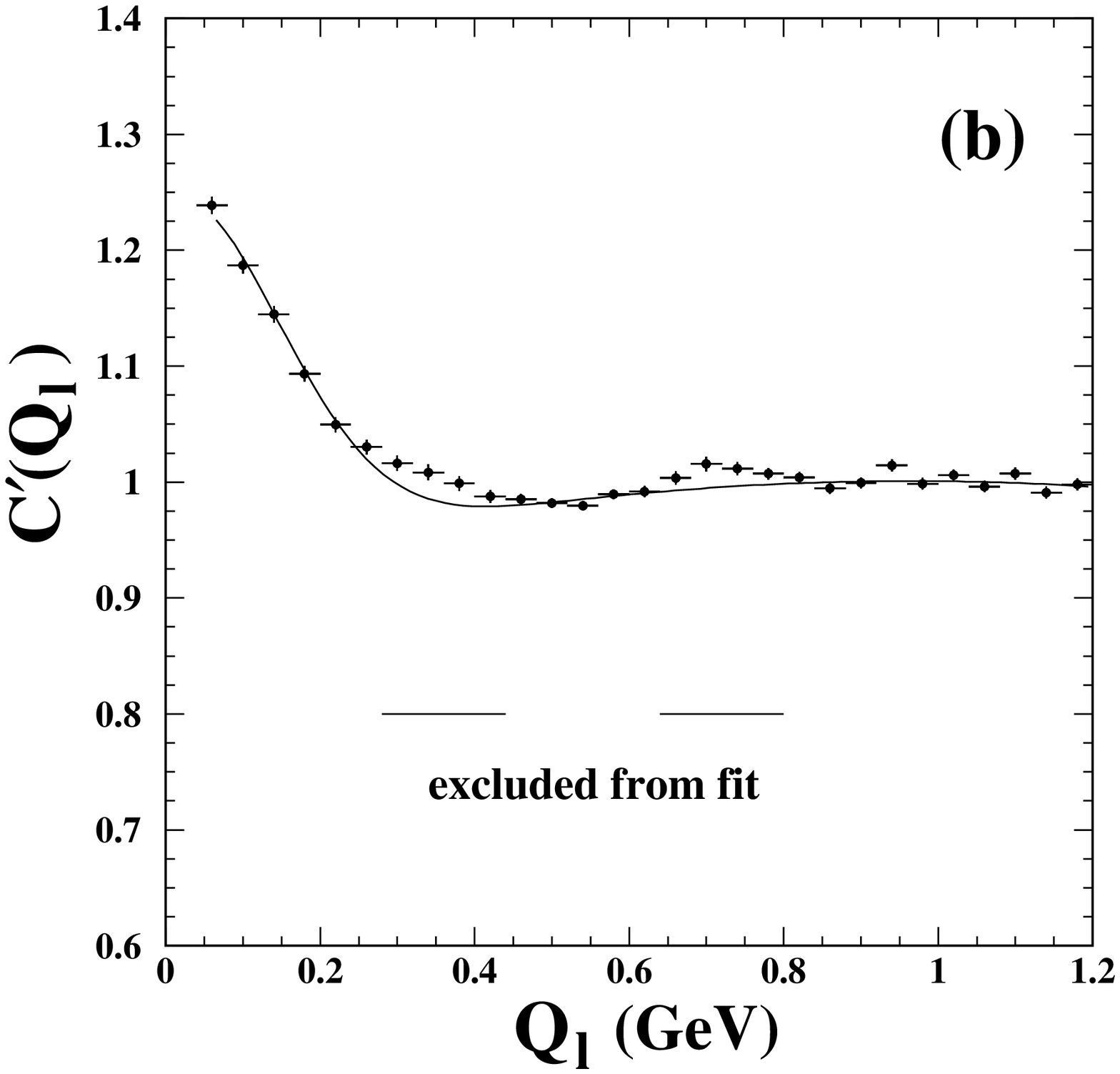}
\end{center}
\caption{Projections of the three-dimensional two-pion correlation function as
  measured by OPAL\protect\cite{2d-opal} (``$+-$'' reference sample).}
\label{fig:2d-opal}
\end{figure}

Recently, it was suggested\cite{markus} that in the framework of
the string model,
the longitudinal source size ought to be larger
than the transverse one. This is now confirmed experimentally by
several LEP experiments\cite{2d-l3,2d-delphi,2d-opal} (see, for
example, the result by OPAL\cite{2d-opal} in
Fig.\ref{fig:2d-opal}). 

Table~\ref{tab:2dim} summarises the measured
values. It should be noted that, in spite of different event selection
criteria and reference samples, all results consistently demonstrate
an elongated shape of the pion source in hadronic $\rZ^0$ decays.

\renewcommand{\arraystretch}{1.3}
\begin{table}[ht]
\vspace*{-0.5cm}
\caption{Elongation of the pion source in hadronic $Z^0$ decays:
  summary of the measurements at LEP1 (\protect$R_{\perp}$ corresponds to
  \protect $Q_{\perp}=\sqrt{Q_{\rt,\out}^2+Q_{\rt,\side}^2}$).\label{tab:2dim}}
\begin{center}
\footnotesize
\begin{tabular}{|l|c|c|c|}
\hline
\begin{minipage}{1cm}
\begin{center}
$\ $\\
$\ $\\
$\ $\\
$\ $
\end{center}
\end{minipage}
&\begin{minipage}{2.5cm}
\begin{center}
L3\\
``mixed'' reference, all events
\end{center}
\end{minipage}
&\begin{minipage}{2.5cm}
\begin{center}
DELPHI\\
``mixed'' reference, 2-jet events
\end{center}
\end{minipage}
&\begin{minipage}{2.3cm}
\begin{center}
OPAL\\
``$+-$'' reference, 2-jet events
\end{center}
\end{minipage}
\\ \hline
$\lambda$&$0.41\pm 0.01^{+0.020}_{-0.019}$&$0.261\pm 0.007\pm 0.010$&$0.443\pm 0.005$\\ \hline
$R_{\lon}$, fm&$0.74\pm 0.02^{+0.04}_{-0.03}$&$0.85\pm 0.02\pm 0.03$&$0.989\pm 0.011^{+0.030}_{-0.015}$\\ \hline
$R_{\rt,\out}$, fm&$0.53\pm 0.02^{+0.05}_{-0.06}$&&$0.647\pm 0.011^{+0.024}_{-0.124}$\\ \hline
$R_{\rt,\side}$, fm&$0.59\pm 0.01^{+0.03}_{-0.13}$&&$0.809\pm 0.009^{+0.019}_{-0.032}$\\ \hline
$R_{\perp}$, fm&&$0.53\pm 0.02\pm 0.03$&\\ \hline
$R_{\perp}/R_{\lon}$&$0.80\pm 0.02^{+0.03}_{-0.18}$&$0.62\pm 0.02\pm 0.04$&$0.818\pm 0.018^{+0.008}_{-0.050}$\\ \hline
\end{tabular}
\end{center}
\vspace*{-0.5cm}
\end{table}

\section{Correlations in kaon pairs}
\label{sec:kaons}
While pion correlations have been extensively studied at LEP,\cite{lepBE}
analysis of BEC in kaon pairs requires a reliable particle
identification, thus very few results were available until now. 
It was
suggested that the observed boson source size should decrease with
increasing particle mass,\cite{alexander} which implies smaller radius
parameter $R$ in case of kaon correlations, as compared to pion. This
hypothesis was recently checked by OPAL:\cite{k-opal} the measured
radius parameter of $R=0.56\pm 0.11$~fm is found to be in the same
range as the one reported earlier by DELPHI.\cite{k-delphi} 
\renewcommand{\arraystretch}{1}

\begin{table}[hb]
\vspace*{-0.5cm}
\caption{Summary of kaon and pion BEC measurements by LEP experiments (charged pairs, one-dimensional $C_2(Q)$).\label{tab:kaon-pion}}
\footnotesize
\begin{center}
\begin{tabular}{|l|c|c|c|c|}
\hline
&\multicolumn{2}{|c|}{``$+-$'' ref.}&
\multicolumn{2}{|c|}{``mixed'' ref.}\\ \cline{2-5}
Experiment&$\lambda$&$R$, fm&$\lambda$&$R$, fm\\
\hline\hline
OPAL, KK, 2-jet&&&0.82$\pm$0.28&0.56$\pm$0.11\\ \hline
DELPHI, KK, all&0.82$\pm$0.27&0.48$\pm$0.08&&\\ \hline \hline
ALEPH, $\pi\pi$, 2-jet&0.48$\pm$0.03&0.81$\pm$0.04&0.30$\pm$0.01&0.51$\pm$0.02\\ \hline
DELPHI, $\pi\pi$, 2-jet&0.31$\pm$0.02&0.83$\pm$0.03&0.24$\pm$0.02&0.47$\pm$0.03\\ \hline
L3, $\pi\pi$, all&0.30$\pm$0.01&0.94$\pm$0.04&0.20$\pm$0.02&0.58$\pm$0.05\\ \hline
OPAL, $\pi\pi$, 2-jet&0.57$\pm$0.04&1.00$\pm$0.10&&\\
\hline
\end{tabular}
\end{center}
\end{table}
As can be
seen in Table~\ref{tab:kaon-pion}, the kaon source size is the same as
for pions, as long as the ``mixed'' reference samples are used for the
latter. However, it is clear that usage of the unlike-sign ``$+-$''
reference systematically leads to a nearly twice bigger observed pion
source sizes. This can be explained by the fact that the residual BEC
in ``$+-$'' pion pairs are much more significant than in kaon pairs.
Thus, this reference is less reliable.

\section{Fermi-Dirac correlations}
\label{sec:fermions}
Two-particle correlations between fermions should exhibit effects of
the Fermi-Dirac statistics. Indirect studies\cite{lambda-spin}
concerning spin composition of $\Lambda\Lambda$ and
$\bar{\Lambda}\bar{\Lambda}$ pairs lead to the observation of a suppression
of S=1 state, and gave estimates of the $R$-parameter being around
$0.1\div 0.2$ fm (see Table~\ref{tab:lambda}). 
\renewcommand{\arraystretch}{1.3}

\begin{figure}[hb]
\begin{center}
\epsfxsize=7cm
\epsfbox{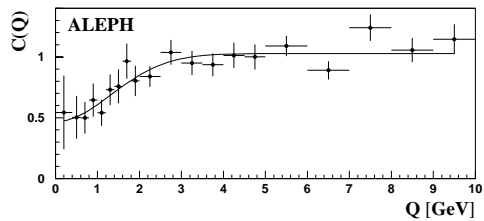}
\end{center}
\caption{Correlation between \protect$\Lambda$ pairs as measured by
  ALEPH\protect\cite{lambda-aleph} (``mixed'' reference).}
\label{fig:lambda-aleph}
\vspace*{-1cm}
\end{figure}

\begin{table}[ht]
\vspace*{-0.5cm}
\caption{\protect$\Lambda$ source size $R$: comparison of different studies .\label{tab:lambda}}
\begin{center}
\footnotesize
\begin{tabular}{|c|l|c|}
\hline
\multicolumn{2}{|l|}{Direct $C_2(Q)$, ALEPH}&$0.11\pm 0.02\pm 0.01$ fm\\ \hline
Spin&ALEPH&$0.14\pm 0.09\pm 0.03$ fm\\ \cline{2-3}
composition&OPAL&$0.14^{+0.37}_{-0.04}$ fm\\ \cline{2-3}
analysis&DELPHI&$0.11^{+0.05}_{-0.03}$ fm\\ \hline
\end{tabular}
\end{center}
\end{table}
Recent analysis by
ALEPH\cite{lambda-aleph} involves direct correlation function
measurement in $\Lambda\Lambda$ and $\bar{\Lambda}\bar{\Lambda}$
pairs, using an inclusive event sample and the ``mixed'' reference sample
as the primary choice. Fig.~\ref{fig:lambda-aleph} clearly shows
the presence of a depletion in $\Lambda$ pair production at small $Q$.

It can be seen in Table~\ref{tab:lambda}, that different methods lead
to rather similar results, which show that the $\Lambda$ source size
appears to be much smaller than that of pions and kaons. 

\section{$m_T$ dependence of the BEC function parameters}
\label{sec:string}
The issue of the transverse mass dependence of the BEC
function parameters in hadronic $\rZ^0$ decays was addressed in
several studies.\cite{os-delphi,os-turin} The fact that the radius
parameters of the two-particle correlation function tend to decrease
with increasing transverse mass is rather astonishing, since a very
similar tendency is observed in heavy-ion collisions.\footnote{see,
  e.g., the talk by M.~Murray in this volume}

\begin{figure}[ht]
\begin{center}
\epsfxsize=5cm
\epsfbox{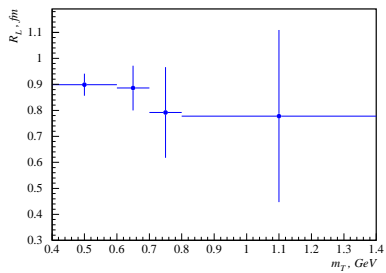}
\end{center}
\caption{Estimated longitudinal radius dependence on $m_T$ for kaon
  pairs in \textsc{Jetset}\protect\cite{os-turin}.}
\label{fig:string}
\end{figure}

It has been long advocated that the proper quantum-mechanical
approach to BEC, involving symmetrization of production amplitudes,
can be introduced into a model like \textsc{Jetset}.  Technically,
it is quite a complicated task, if all the bosons have to be considered. It
is feasible though to implement such a procedure for the case of
charged kaons,\cite{os-turin} since normally there is only one pair of
identical charged kaons in a $\rZ^0$ decay event. It was shown that,
upon applying the symmetrization, kaon pairs exhibit a characteristic
Bose-Einstein enhancement.\cite{os-turin} Moreover, as can be seen in
Fig.~\ref{fig:string}, the symmetrization effectively causes kaons to
be produced closer to each other, if their transverse mass increases. 
Whether caused by the explicit $m_T$ dependence of the fragmentation function,
or by the local energy-momentum conservation, this effect is likely to
be the same as the one observed in two-pion correlations.

\section{Summary}
\label{sec:summary}
Analysis of LEP1 data produced many important results for
multiparticle correlation studies. Among the most recent results is
the measurement of the pion source elongation by different experiments,
which found a longitudinal radius exceeding the transverse ones by 25\%
to 60\%. Measurement of the BEC parameters for charged kaons showed that
they are similar to those of pions; no strong evidence for a smaller
radius exists at the moment. Correlations between $\Lambda$ baryons
were measured directly, and the observed source size was confirmed to
be much smaller than that of bosons. Studies of the $m_\rT$ dependence
of the BEC function parameters indicate that, like in heavy-ion collisions,
the observed source size tends to
decrease with increasing $m_\rT$ of the boson pairs. 
This effect is to big extent reproduced by \textsc{Jetset},
and attempts were done to explain it in the framework of the string model.

\end{document}